\documentclass[nofootinbib,preprint,superscriptaddress,showpacs,amsmath,amssymb,aps,pra,showkeys]{revtex4-1}

\usepackage{amsmath,amsfonts,amssymb}
\usepackage{graphicx}
\usepackage{color}

\usepackage{calc}

\usepackage{accents}
\newcommand{\dbtilde}[1]{\accentset{\approx}{#1}}

\begin{document}

\title{Semiclassical two-step model with quantum input: Quantum-classical approach to strong-field ionization}

\author{N. I. Shvetsov-Shilovski}
\email{n79@narod.ru}
\affiliation{Institut f\"{u}r Theoretische Physik, Leibniz Universit\"{a}t Hannover, D-30167 Hannover, Germany}

\author{M. Lein}
\affiliation{Institut f\"{u}r Theoretische Physik, Leibniz Universit\"{a}t Hannover, D-30167 Hannover, Germany}

% \author{L. B. Madsen}
% \affiliation{Department of Physics and Astronomy, Aarhus University, 8000 {\AA}rhus C, Denmark}

\date{\today}

\begin{abstract}
We present a mixed quantum-classical approach to strong-field ionization - a semiclassical two-step model with quantum input. In this model the initial conditions for classical trajectories that simulate electron wave packet after ionization are determined by the exact quantum dynamics. As a result, the model allows to overcome deficiencies of standard semiclassical approaches in describing the ionization step. The comparison with the exact numerical solution of the time-dependent Schr\"{o}dinger equation shows that for ionization of a one-dimensional atom the model yields quantitative agreement with the quantum result. This applies both to the width of the photoelectron momentum distribution and the interference structure.\\

% \noindent Discipline: Atomic, Molecular and Optical, Facet: Research Areas, Concept: Multiphoton or tunneling ionization and excitation.\\
% Facet: Physical Systems, Concept: Atoms.\\
% Facet: Techniques, Concepts: Semiclassical methods, Strong-field approximation. 
 
\end{abstract}

% \pacs{32.80Fb, 32.80Rm, 32.80 Wr}

\maketitle

\section{Introduction} Strong-field physics is a fascinating field of research resulting from remarkable progress in laser technologies during the last three decades. The interaction of strong laser radiation with atoms and molecules leads to many highly nonlinear phenomena, including above-threshold ionization (ATI) along with the formation of the plateau in the energy spectrum of the photoelectrons (high-order ATI), generation of high-order harmonics of the incident field (HHG), and nonsequential double ionization (NSDI) (see, e.g., Refs.~\cite{DeloneKrainovBook2000, BeckerRev2002, MilosevicRev2003, FaisalRev2005, FariaRev2011} for reviews). The main theoretical approaches used to study all these phenomena are based on the strong-field approximation (SFA) \cite{Keldysh1964, Faisal1973, Reiss1980}, direct numerical solution of the time-dependent Schr\"{o}dinger equation (TDSE) (see, e.g, Refs. \cite{Muller1999, Bauer2006, Madsen2007, Grum2010, Patchkovskii2016, Tong2017} and references therein), and the semiclassical models. 

The semiclassical models apply classical mechanics to describe the motion of an electron after it has been released from an atom or molecule by a strong laser field. The most widely known examples of the semiclassical approaches are the two-step \cite{Linden1988, Gallagher1988, Corkum1989} and the three-step \cite{Kulander_Schafer1993, Corkum1993} models. The two-step model corresponds to the following picture of ionization process. In the first step an electron is promoted into the continuum, typically by tunneling ionization \cite{Dau, PPT, ADK}. In the second step the electron moves in the laser field towards a detector along a classical trajectory. In addition to these two steps, the three-step model involves the interaction of the returning electron with the parent ion. Accounting for this interaction allows the three-step model to qualitatively describe high-order ATI, HHG, and NSDI. 

The semiclassical approaches have important advantages. First, the trajectory-based models, including those that take into account both the laser field and the ionic potential, are often computationally simpler than the numerical solution of the TDSE. What is even more important, the analysis of the classical trajectories helps to understand the physical picture of the strong-field phenomenon under study. 

In order to calculate the classical trajectory, it is necessary to specify the corresponding initial conditions, i.e., the starting point and the initial velocity of the electron. To obtain the former, i.e., the tunnel exit point, the separation of the tunneling problem for the Coulomb potential in parabolic coordinates can be used, see, e.g., Ref.~\cite{Dau}. In trajectory-based models it is often assumed that the electron starts with zero initial velocity along the laser field. Simultaneously, it can have a nonzero initial velocity in the direction perpendicular to the field. The initial transverse momenta, as well as the instants of ionization, are usually distributed in accord with the static ionization rate \cite{DeloneKrainov1991} with the field strength equal to the instantaneous field at the time of ionization.     

In the standard formulation, the trajectory models used in strong-field physics are not able to describe quantum interference effects. Accounting for interference effects in trajectory-based simulations have attracted considerable interest (see, e.g., Refs.~\cite{SandRost2000, Spanner2003, Zagoya2012, Faria2014}). The recently developed quantum trajectory Monte-Carlo (QTMC) \cite{Li2016} and semiclassical two-step  (SCTS) models \cite{Shvetsov2016} describe interference structures in photoelectron momentum distributions of the ATI process. These models assign a certain phase to each classical trajectory, and the corresponding contributions of all the trajectories leading to a given asymptotic (final) momentum are added coherently. The QTMC model accounts for the Coulomb potential within the semiclassical perturbation theory. In contrast to this, in the SCTS the phase associated with every trajectory is obtained using the semiclassical expression for the matrix element of the quantum-mechanical propagator (see Ref.~\cite{Miller1974}). Therefore, the SCTS model accounts for the binding potential beyond the semiclassical perturbation theory. This explains why for identical initial conditions after the ionization step the SCTS model shows closer agreement with solution of the TDSE than the QTMC model (see Ref.~\cite{Shvetsov2016}).  

The analysis of the photoelectron momentum distributions and energy spectra calculated within both the QTMC and the SCTS models showed that the ATI peaks are qualitatively reproduced by the semiclassical approaches \cite{Shvetsov2016}. However, the semiclassical approximation does not quantitatively reproduce the amplitude of the oscillations. The photoelectron spectra calculated within the semiclassical models fall off too rapidly for energies exceeding $U_{p}$, where $U_p=F_{0}^{2}/4\omega^2$ is the ponderomotive energy, i.e., the cycle-averaged quiver energy of a free electron in an electromagnetic field (atomic units are used throughout the paper unless indicated otherwise). Here, $F_{0}$ and $\omega$ are the amplitude and the frequency of the field, respectively. This deficiency is closely related to the fact that the initial conditions usually employed in semiclassical models provide too few trajectories with large longitudinal momenta \cite{Shvetsov2016}.    

Recently several approaches to improving the quality of the initial conditions in semiclassical models have been proposed. The simplest method is to distribute the initial conditions for electron trajectories using the SFA formulas, see, e.g., Refs.~\cite{Yan2010, Popruzhenko2008, Boge2013, Hofmann2014, Geng2014, Li2016dop}. We note that this method dates back to Refs.~\cite{Yudin2001, Bondar2009}. In most cases it leads to closer agreement with the TDSE. However, to the best of our knowledge, the validity of the usage of the SFA expressions in trajectory-based simulations has not been systematically analyzed so far. In this paper we consider an alternative approach: the combination of the semiclassical models with direct numerical solution of the TDSE. 

A significant step in this direction has been taken with the development of the backpropagation method (see Refs.~\cite{Ni2016, Ni2018}). In this method, the wave packet of the outgoing electron obtained from the TDSE is transformed into classical trajectories. These trajectories are then propagated backwards in time, which makes it possible to retrieve the information about the tunnel exit point and the initial electron velocity. Various approximations to the distributions of the starting points and the initial velocities were analyzed by choosing different criteria to stop the backpropagating trajectories \cite{Ni2018, Niwe2018}. However, the backpropagation method requires the numerical solution of the TDSE up to some point in time after the end of the laser pulse \textit{in the whole space}. This restricts its applicability in the case of computationally difficult strong-field problems.  

A promising approach would be a combination of the SCTS model with extended virtual detector theory (EVDT), see Refs.~\cite{Wang2013, Wang2018}. For the first time the concept of virtual detector (VD) was proposed in Ref.~\cite{Thumm2003} as a method for calculating momentum distributions from the time-dependent wave function. The EVDT approach combines the VD method with semiclassical simulations. The EVDT employs a network of virtual detectors that encloses an atom interacting with the external laser field. Each detector detects the wave function $\psi\left(\vec{r},t\right)=A\left(\vec{r},t\right)\exp\left[i\phi\left(\vec{r},t\right)\right]$ obtained by solving the TDSE and generates a classical trajectory at the same position with the initial momentum $\vec{k}$ determined from the gradient of the phase, $\vec{k}\left(\vec{r}_{d},t\right)\equiv\nabla\cdot\phi\left(\vec{r}_{d},t\right)=\vec{j}\left(\vec{r}_{d},t\right)/\left|A\left(\vec{r}_{d},t\right)\right|^{2}$. Here $\vec{r}_{d}$ is the position of a virtual detector and $\vec{j}\left(\vec{r}_{d},t\right)$ is the probability flux at this position. The latter determines the relative weight of the generated trajectory. The subsequent motion of an electron is found from the solution of Newton's equations. The final photoelectron momentum distribution is obtained by summing over all classical trajectories with their relative weights. It should be stressed that EVDT solves the TDSE only within some restricted region centered at the atom. A network of virtual detectors is placed at the boundary of this region. This reduces the computational load of numerically difficult strong-field problems. Recently the VD approach was used for study of tunneling times \cite{Teeny2016} and longitudinal momentum distributions \cite{Tian2017} in strong-field ionization. 

Leaving the combination of the SCTS with the EVDT for future studies, in this paper we formulate an alternative quantum-classical approach: the semiclassical two-step model with quantum input (SCTSQI). To this end, we combine the SCTS model with initial conditions obtained from TDSE solutions using Gabor transforms. For simplicity, we consider ionization of a one-dimensional (1D) atom. The generalization to the real three-dimensional case is straightforward. The benefit of the 1D model, however, is that potential deficiencies of trajectory models are exposed better and, therefore, it makes the comparison with the fully quantum simulations more valuable. 

The paper is organized as follows. In Sec.~II we sketch our approach to solve the TDSE, we briefly review the SCTS model, and we formulate our SCTSQI approach. In Sec.~III we apply our model to the ionization of a 1D model atom and present comparison with the TDSE results. The conclusions and outlook are given in Sec.~IV.   
		
\section{Semiclassical two-step model with quantum input}
We benchmark our semiclassical model against the results obtained by direct numerical solution of the 1D TDSE and by using the SCTS model. For this reason, before formulating the SCTSQI model and discussing its outcomes, we briefly review the technique used to solve the TDSE and sketch the SCTS model. We define a few-cycle laser pulse linearly polarized along the $x$-axis in terms of a vector-potential:
\begin{equation}
\vec{A}=\left(-1\right)^{n+1}\frac{F_{0}}{\omega}\sin^2\left(\frac{\omega t}{2n}\right)\sin\left(\omega t +\varphi\right)\vec{e}_{x}.
\label{vecpot}
\end{equation} 
Here $n$ is the number of the cycles within the pulse, $\varphi$ is the carrier envelope phase, and $\vec{e}_{x}$ is a unit vector. The laser pulse is present between $t=0$ and $t_f=\left(2\pi/\omega\right)n$, and its electric field $\vec{F}$ can be obtained from Eq.~(\ref{vecpot}) by $\vec{F}=-\frac{d\vec{A}}{dt}$. 

\subsection{Solution of the one-dimensional time-dependent Schr\"{o}dinger equation}
In the velocity gauge, the 1D TDSE for an electron in the laser pulse reads as:
\begin{equation}
i\frac{\partial}{\partial t}\Psi\left(x,t\right)=\left\{\frac{1}{2}\left(-i\frac{\partial}{\partial x}+A_{x}\left(t\right)\right)^2+V\left(x\right)\right\}\Psi\left(x,t\right),
\label{tdse}
\end{equation}
where $\Psi\left(x,t\right)$ is the time-dependent wave function in coordinate space, and  a soft-core potential $V\left(x\right)=-\frac{1}{\sqrt{x^2+a^2}}$ is used, with $a=1$ as in Ref.~\cite{Javanainen1988}. 

In the absence of the laser pulse, the 1D system satisfies the time-independent Schr\"{o}dinger equation:
\begin{equation}
\left\{-\frac{1}{2}\frac{d^2}{dx^2}+V\left(x\right)\right\}\Psi\left(x\right)=E\Psi\left(x\right).
\label{tise}
\end{equation} 
We solve Eq.~(\ref{tise}) on a grid and approximate the second derivative by the well-known three-point formula. For our simulations we use a box centered at the origin and extending to $\pm x_{\text{max}}$, i.e., $x\in\left[-x_{\text{max}},x_{\text{max}}\right]$. Typically, our grid extends up to $x_{\text{max}}=500$~a.u. and consists of 8192 points, which corresponds to the grid spacing $dx\approx0.1221$~a.u. The energy eigenvalues $E_{n}$ and the corresponding eigenfunctions $\Psi_{n}\left(x\right)$ are found by a diagonalization routine designed for sparse matrices \cite{Lapack}. For the chosen value of $a$ we find the ground-state energy $E_{0}=-0.6698$~a.u. This value, as well as the energies of other lowest-energy bound states, coincide with the results of Ref.~\cite{Javanainen1988}.

We solve Eq.~(\ref{tdse}) using the split-operator method \cite{Feit1982} with the time step $\Delta t=0.0734$~a.u. Unphysical reflections of the wave function from the grid boundary is prevented by using absorbing boundaries. More specifically, in the region $\left|x\right|\geq x_{b}$ we multiply the wave function by a mask 
\begin{equation}
M\left(x\right)=\cos^{1/6}\left[\frac{\pi\left(\left|x\right|-x_{b}\right)}{2\left(x_{\text{max}}-x_{b}\right)}\right].
\label{mask}
\end{equation}
Here we assume that the internal boundaries of the absorbing regions correspond to $x=\pm x_{b}$ (we use $x_b=3x_{\text{max}}/4$). This ensures that the part of the wave function in the mask region is absorbed without an effect on the inner part $\left|x\right|<x_{b}$. We calculate the photoelectron momentum distributions using the mask method (see Ref.~\cite{Tong2006}).  

\subsection{Semiclassical two-step model}
In our semiclassical simulations the trajectory $\vec{r}\left(t\right)$ and momentum $\vec{p}\left(t\right)$ of an electron are calculated treating the electric field of the pulse $\vec{F}\left(t\right)$ and the ionic potential $V\left(\vec{r},t\right)$ on equal footing:
\begin{equation}
\frac{d^{2}\vec{r}}{dt^2}=-\vec{F}\left(t\right)-\vec{\nabla}V\left(\vec{r},t\right).
\label{newton}
\end{equation}
In the SCTS, every trajectory is associated with the phase of the matrix element of the semiclassical propagator \cite{Miller1974}. For an arbitrary effective potential $V\left(\vec{r},t\right)$ the SCTS phase reads as:
\begin{equation}
\Phi\left(t_{0},\vec v_0\right)= - \vec v_0\cdot\vec r(t_0) + I_{p}t_{0} -\int_{t_0}^\infty dt\ \left\lbrace\frac{p^2(t)}{2}+V[\vec{r}(t)]-\vec r(t)\cdot\vec\nabla V[\vec{r}(t)]\right\rbrace\,
\label{Phi_sim}
\end{equation}
where $t_0$ is the ionization time, and $\vec{r}\left(t_0\right)$ and $\vec{v}_{0}$ are the initial electron position and velocity of an electron, respectively. 

In the importance sampling implementation of the SCTS model the ionization times $t_0^{j}$ and transverse initial velocities $\vec{v}_{0,\perp}^{j}$ $\left(j=1,...,n_p\right)$ of the ensemble consisting of $n_p$ trajectories are distributed in accord with the square root of the tunneling probability (see Ref.~\cite{Shvetsov2016}). The latter is given by the formula for the static ionization rate \cite{DeloneKrainov1991}:
\begin{equation}  
w\left(t_{0},v_{0, \perp}\right)\sim\exp\left(-\frac{2\left(2I_p\right)^{3/2}}{3F\left(t_0\right)}\right)\exp\left(-\frac{\kappa v_{0,\perp}^{2}}{F\left(t_0\right)}\right),
\label{tunrate}
\end{equation}
where $I_p$ is the ionization potential and $\vec{v}_{0,\perp}$ is the initial velocity in the direction perpendicular to the laser field. We solve Newton's equations of motion (\ref{newton}), in order to find the final (asymptotic) momenta of all the trajectories, and bin them in cells in momentum space according to these final momenta. The contributions of the $n_{\vec{k}}$ trajectories that reach the same bin centered at a given final momentum $\vec{k}$ are added coherently, and, as the result, the ionization probability $R(\vec{k})$ is given by:
\begin{equation}
\label{prob}
R(\vec{k})=\left|\sum_{j=1}^{n_{\vec{k}}}\exp\left[i\Phi\left(t_{0}^{j},\vec v_0^{j}\right)\right]\right|^{2}.
\end{equation} 

We note that the application of the importance sampling technique is not the only possible way to implement the SCTS model: The initial conditions can be distributed either randomly or, alternatively, a uniform grid in the $\left(t_0,\vec{v}_{0}\right)$ space can be used. In both latter cases Eq.~(\ref{prob}) is replaced by:
\begin{equation}
\label{prob2}
R(\vec{k})=\left|\sum_{j=1}^{n_{\vec{k}}}\sqrt{w\left(t_{0}^{j},\vec v_0^{j}\right)}\exp\left[i\Phi\left(t_{0}^{j},\vec v_0^{j}\right)\right]\right|^{2},
\end{equation}
In the present work, we use random distributions of $t_0$ and $\vec{v}_{0}$. 

If the potential $V\left(\vec{r},t\right)$ is set to the 1D soft-core potential $V\left(x\right)=-1/\sqrt{x^2+a^2}$, the equation of motion (\ref{newton}) and the expression for the SCTS phase (\ref{Phi_sim}) reads as
\begin{equation}
\frac{d^{2}x}{dt^2}=-{F}_{x}\left(t\right)-\frac{x}{\left(x^2+a^2\right)^{3/2}},
\label{newton_1d}
\end{equation}
and, choosing the initial velocity as zero, we have the phase
\begin{equation}
\Phi\left(t_{0},\vec{v}_{0}\right)= I_{p}t_{0} -\int_{t_0}^\infty dt \left\{\frac{v_{x}^{2}\left(t\right)}{2}-\frac{x^2}{\left(x^2+a^2\right)^{3/2}}-\frac{1}{\sqrt{x^2+a^2}}\right\}dt.
\label{Phi_sim_1d}
\end{equation}
In the 1D case the ionization rate (\ref{tunrate}) is replaced by 
\begin{equation}
w\left(t_0\right)\sim\exp\left(-\frac{2\left(2\left|E_{0}\right|\right)^{3/2}}{3F\left(t_0\right)}\right),
\end{equation}
where $E_0=-0.6698$~a.u. is the ground-state energy in the potential $V\left(x\right)$. 

We integrate the equation of motion numerically up to $t=t_{f}$ and find the final electron momentum $k_x$ from its momentum $p_x\left(t_f\right)$ and position $x\left(t_f\right)$ at the end of the laser pulse. To this end, the energy conservation law can be used. Since an unbound classical electron cannot change the direction of its motion at $t \geq t_f$, the sign of the $k_x$ coincides with that of $p_x\left(t_f\right)$.

In order to accomplish the formulation of the SCTS model for the 1D case, we need to calculate the post-pulse phase, i.e., the contribution to the phase (\ref{Phi_sim_1d}) accumulated in the asymptotic interval $\left[t_f,\infty\right]$. Indeed the phase of Eq.~(\ref{Phi_sim_1d}) can be decomposed as:
\begin{equation}
\Phi\left(t_{0},\vec{v}_{0}\right)= I_{p}t_{0} -\int_{t_0}^{t_f} dt \left\{\frac{v_{x}^{2}\left(t\right)}{2}-\frac{x^2}{\left(x^2+a^2\right)^{3/2}}-\frac{1}{\sqrt{x^2+a^2}}\right\}+\Phi_{f}^{V},
\label{decomp}
\end{equation} 
where the post-pulse phase $\Phi_{f}^{V}$ reads
\begin{equation}
\Phi_{f}^{V}\left(t_f\right)=-\int_{t_f}^{\infty}\left(E-\frac{x^2\left(t\right)}{\left[x^2\left(t\right)+a^2\right]^{3/2}}\right)dt
\label{postpulse}
\end{equation} 
with total energy $E$. As in Ref.~\cite{Shvetsov2016}, we separate the phase (\ref{postpulse}) into parts with time-independent and time-dependent integrand. The first part yields the linearly divergent contribution 
\begin{equation}
\lim_{t\to\infty}(t_f-t)E
\end{equation} 
that is to be disregarded, since it results to the zero phase difference for the trajectories leading to the same momentum cell. Therefore, the post-pulse phase is determined by the time-dependent contribution
\begin{equation}
\tilde{\Phi}_{f}^{V}=\int_{t_f}^{\infty}\frac{x^2\left(t\right)}{\left[x^2\left(t\right)+a^2\right]^{3/2}}dt.
\label{postpulse2}
\end{equation}
Although the integral (\ref{postpulse2}) diverges, we can isolate the divergent part as follows:
\begin{equation}
\tilde{\Phi}_{f}^{V}=\int_{t_f}^{\infty}\left[\frac{x^{2}}{\left(x^2+a^2\right)^{3/2}}-\frac{2Et^{2}}{\left(2Et^2+a^2\right)^{3/2}}\right]dt+\int_{t_f}^{\infty}\frac{2Et^{2}}{\left(2Et^2+a^2\right)^{3/2}}dt.
\label{regul}
\end{equation}
The divergent contribution, i.e., the second term of Eq.~(\ref{regul}), depends only on the electron energy $E$ and parameter $a$ and, therefore, is equal for all the trajectories leading to a given bin on the $p_x$ axis. Since we are interested in the relative phases of the interfering trajectories, this common divergent part can be omitted, and the post-pulse phase can be calculated as
\begin{equation}
\dbtilde{\Phi}_{f}^{V}=\int_{t_f}^{\infty}\left[\frac{x^{2}}{\left(x^2+a^2\right)^{3/2}}-\frac{2Et^{2}}{\left(2Et^2+a^2\right)^{3/2}}\right]dt.
\label{post_pulse_final}  
\end{equation}
The integral in Eq.~(\ref{post_pulse_final}) converges and can be easily calculated numerically. It depends on the electron position $x\left(t_f\right)$ and velocity $v_x\left(t_f\right)$ at the end of the pulse. In practice, we calculate this integral on a grid in the $\left(x\left(t_f\right),v_{x}\left(t_f\right)\right)$ plane and use bilinear interpolation, in order to find its value for $x\left(t_f\right)$ and $v_x\left(t_f\right)$ that correspond to every electron trajectory.

\subsection{Semiclassical two-step model with quantum input}
Combination of the exact solution of the TDSE with a trajectory-based model is not a simple task. In order to calculate a classical trajectory, both the starting point and the initial velocity are needed. However, in accord with Heisenberg's uncertainty principle, there is a fundamental limit to the precision with which canonically conjugate variables as position and momentum can be known. % The time-dependent wave function $\Psi\left(x,t\right)$ in coordinate representation obtained from the TDSE (\ref{tdse}) does not contain any information regarding electron momentum. Similarly, the Fourier transform of $\Psi\left(x,t\right)$ in the whole space, i.e., 
% \begin{equation}
% \phi\left(p_x,t\right)=\frac{1}{\sqrt{2\pi}}\int_{-\infty}^{\infty}\Psi\left(x,t\right)\exp\left(-ip_xx\right)dx
% \end{equation}
% cannot be used to describe the probability of finding a quantum particle in a certain point of the coordinate space. 
Information about \textit{both} the position and momentum of a quantum particle can be obtained using a position-momentum quasiprobability distribution, e.g., the Wigner function or Husimi distribution (see, e.g., Ref.~\cite{Ballentine} for a text-book treatment). Here we employ the Gabor transformation \cite{Gabor}, which is widely used for the analysis of the HHG and ATI, see, e.g., Refs.~\cite{Chirila2010, Bandrauk2012, Wu2013, Shu2016}. The Gabor transformation of a function $\tilde{\Psi}\left(x,t\right)$ near the point $x_0$ is defined by:
\begin{equation}
G\left(x_0,p_x,t\right)=\frac{1}{\sqrt{2\pi}}\int_{-\infty}^{\infty}\tilde{\Psi}\left(x^{\prime},t\right)\exp\left[-\frac{\left(x^{\prime}-x_{0}\right)^2}{2\delta_{0}^{2}}\right]\exp\left(-ip_xx^{\prime}\right)dx^{\prime},
\label{Gabor}
\end{equation}
where the exponential factor $\exp\left[-\frac{\left(x^{\prime}-x_{0}\right)^2}{2\delta_{0}^{2}}\right]$ is a window with the width $\delta_{0}$. The squared modulus of $G\left(x_0,p_x,t\right)$ describes the momentum distribution of the electron in the vicinity of $x=x_{0}$ at time $t$. In fact, $\left|G\left(x_0,p_x,t\right)\right|^{2}$ is nothing but the Husimi distribution \cite{Husimi}, which can be obtained by a Gaussian smoothing of the Wigner function. In contrast to the Wigner function, the Husimi distribution is a positive-semidefinite function, which facilitates the interpretation as a quasiprobability distribution. 
In our SCTSQI model, we solve the TDSE in the length gauge:
\begin{equation}
i\frac{\partial}{\partial t}\Psi\left(x,t\right)=\left\{-\frac{1}{2}\frac{\partial^2}{\partial x^2}+V\left(x\right)+F_{x}\left(t\right)x\right\}\Psi\left(x,t\right).
\label{tdse_len}
\end{equation}
We introduce two additional spatial grids consisting of $N$ points in the absorbing regions of the computational box:
\begin{equation}
x_{0,\pm}^{k}=\mp\left(x_{b}+\Delta x \cdot k\right),
\label{grids}
\end{equation}
where $\Delta x=\left(x_{\text{max}}-x_{b}\right)/N$ and $k=1,...,N$. At every step of the time propagation of the TDSE (\ref{tdse_len}) we calculate the Gabor transform (\ref{Gabor}) of the absorbed part $\tilde{\Psi}$ at the points $x_{0,-}^{k}$ and $x_{0,+}^{k}$, see Fig.~1~(a). As a result, at every time instant $t$ we know $G\left(x,p_x,t\right)$ on the grids in the rectangular domains $D_{1}=\left[-x_{\text{max}},-x_{b}\right]\times\left[-p_{x,\text{max}}, p_{x,\text{max}}\right]$ and $D_{2}=\left[x_{b},x_{\text{max}}\right]\times\left[-p_{x,\text{max}}, p_{x,\text{max}}\right]$ of the phase space. Here $p_{\text{max}}$ is the maximum momentum, i.e., $p_{\text{max}}=\pi/\Delta x$, if the fast Fourier transform is used to calculate Eq.~(\ref{Gabor}). 
An example of the corresponding Husimi quasiprobability distribution calculated at $t=3t_f/2$ is shown in Fig.~1~(b). At this time instant the quasiprobability distribution consists of the three main spots $P_1$, $P_2$, and $P_3$, whose maxima are indicated by a (green) circle, (magenta) square and (cyan) triangle, respectively. These maxima correspond to the electron momenta $k_x$ equal to $0.37$~a.u., $-0.17$~a.u., and $-0.48$~a.u., respectively [see Fig.~1(b)]. According to the two-step model, a final electron momentum $k_x$ corresponds to the ionization times $t_0$ satisfying the equation
\begin{equation}
k_x=-A_x\left(t_0\right).
\label{two_step}
\end{equation}
Depending on the momentum value, this equation can have several solutions, and therefore, several different ionization times can lead to a given $k_x$, see Fig.~1~(c), which shows the final electron momentum as a function of the ionization time. The analysis of the time evolution of the electron probability density reveals that every spot in Fig.~1~(b) is mainly created within a narrow time interval that is close to only one of the solutions of Eq.~(\ref{two_step}). The solutions of Eq.~(\ref{two_step}) that make the main contributions to the maxima of $P_1$, $P_2$, and $P_3$ are shown in Fig.~1~(c). This fact is easy to understand, if we take into account that Fig.~1~(b) is a snapshot of the dynamic quasiprobability distribution in the absorbing mask regions. Indeed, at a given time instant the contributions to the Husimi distribution from the vicinities of other solutions of Eq.~(\ref{two_step}) are either already absorbed by the mask, or have not reached the absorbing regions yet. We note that aside from $P_1$, $P_2$, $P_3$ some other less pronounced spots are also seen in Fig.~1~(b). These latter spots correspond to the contributions that by the given time instant are already mostly absorbed. The slight slope of the whole Husimi distribution that is visible in Fig.~1~(b) is due to the fact that the contributions corresponding to the high values of $\left|k_x\right|$ travel larger distances before being absorbed than the ones with smaller $\left|k_x\right|$.
\begin{figure}[h]
\begin{center}
\includegraphics[width=.60\textwidth]{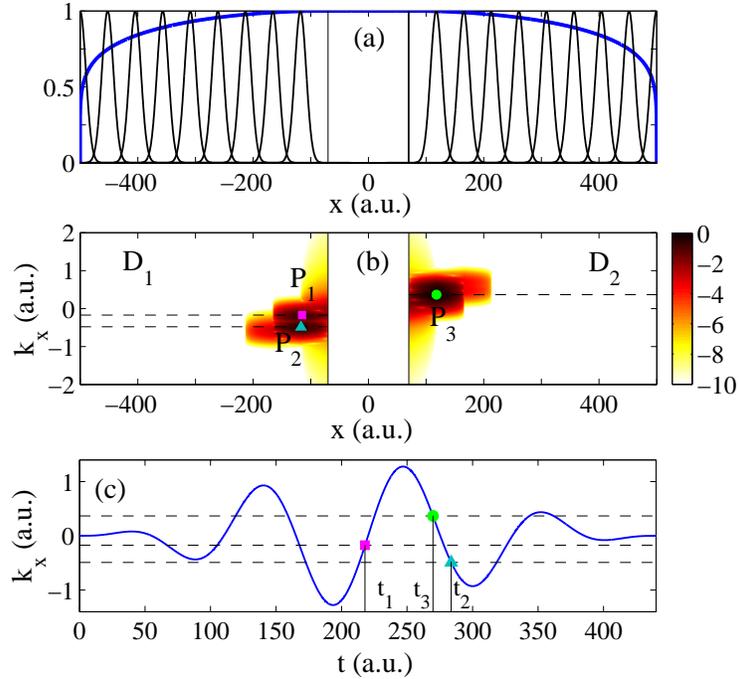} 
\end{center}
\caption{(a) Scheme illustrating the structure of the computational box in the SCTSQI model. The mask function [Eq.~(\ref{mask})] is shown by the blue (thick) curve. The vertical lines correspond to the internal boundaries of the mask region. The black (thin) curves show the windows of the Gabor transform centered at the points $x_{0,\pm}^{i}$ [Eq.~(\ref{grids})]. (b) The Husimi quasiprobability distribution $\left|G\left(x,p_x,t_f/2\right)\right|^{2}$ calculated at $t=3t_f/2$ for the laser pulse defined by Eq.~(\ref{vecpot}) with a duration of $n=4$ cycles, intensity of $2.0\cdot10^{14}$ W/cm$^2$, phase $\varphi=0$, and a wavelength of 800 nm. The Husimi distribution is calculated in the domains $D_1$ and $D_2$ of the phase space (see text). A logarithmic color scale is used. $P_1$-$P_3$ represent the three main spots of the Husimi distribution. The maxima of these spots are depicted by a (green) circle, (magenta) square, and (cyan) rectangle, respectively. (c) The final electron momentum $-A_{x}\left(t\right)$ in the potential-free classical model as a function of the time of ionization. The parameters of the laser pulse are the same as in Fig.~1~(b). The vicinities of the time instants $t_1$, $t_2$, and $t_3$ make the main contribution to the spots $P_1$, $P_2$, and $P_3$, respectively [see Fig.~1~(b)].}
\label{fig2}
\end{figure}

The value of the Gabor transform at an arbitrary point that belongs to the domain $D_1$ or $D_2$ can be obtained by a two-dimensional interpolation. At every time $t_0$ we randomly distribute initial positions $x_{0}^{j}$ and momenta $p_{x,0}^{j}$ $\left(j=1,...,n_{p}\right)$ of $n_p$ classical trajectories in the domains $D_1$ and $D_2$. These trajectories are propagated according to Newton's equation of motion (\ref{newton_1d}). Every trajectory is assigned with the quantum amplitude $G\left(t_0,x_{0}^{j},p_{x,0}^{j}\right)$ and the phase
\begin{equation}
\Phi_{0}\left(t_{0},x_{0}^{j},p_{x,0}^{j}\right)=-\int_{t_0}^{\infty} dt \left\{\frac{v_{x}^{2}\left(t\right)}{2}-\frac{x^2}{\left(x^2+a^2\right)^{3/2}}-\frac{1}{\sqrt{x^2+a^2}}\right\}
\label{phase_sctsqi}
\end{equation}
We note that the SCTSQI phase (\ref{phase_sctsqi}) corresponds to the phase of the matrix element of the semiclassical propagator that describes a transition from momentum $p_{x,0}^{j}$ at $t=t_0$ to momentum $k_{x}^{j}=k_{x}^{j}\left(x_{0}^{j},p_{x,0}^{j}\right)$ at $t\to\infty$. The ionization probability in the SCTSQI is given by
\begin{equation}
R\left(k_x\right)=\left|\sum_{m=1}^{N_{T}}\sum_{j=1}^{n_{k_x}}G\left(t_{0}^{m},x_{0}^{j},p_{x,0}^{j}\right)\exp\left[i\Phi_{0}\left(t_{0}^{m},x_{0}^{j},p_{x,0}^{j}\right)\right]\right|^{2},
\label{sctsqi}
\end{equation} 
where \textcolor{blue}{$N_T$} is the number of the time steps used to solve the TDSE, and $n_{k_x}$ is the number of trajectories reaching the same bin centered at $k_x$ [cf. Eq.~(\ref{prob2})]. It should be stressed that the Gabor transform $G\left(t_{0}^{m},x_{0}^{j},p_{x,0}^{j}\right)$ is a complex function with both absolute value and phase. In order to ensure that ionized parts of the wave function reach the absorbing regions, we propagate the TDSE up to some time $t=T$, where $T>t_f$. For this reason, in the SCTSQI we calculate classical trajectories till $t=T$ and replace $t_f$ by $T$ in Eq.~(\ref{post_pulse_final}) for the post-pulse phase. In our simulations we have used $T=4t_f$.

\section{Results and discussion}
For our numerical examples we use the intensity of $2.01\cdot10^{14}$ W/cm$^2$ ($F_0=0.0757$~a.u.) and the wavelength 800 nm ($\omega=0.057$~a.u.). This corresponds to the Keldysh parameter $\gamma=\omega\sqrt{2I_{p}}/F_{0}$ (see Ref.~\cite{Keldysh1964}) equal to 0.87. For simplicity, we set the absolute phase of the pulse (\ref{vecpot}) equal to zero: $\varphi=0$.   

We benchmark our SCTSQI approach against the SCTS model and the exact numerical solution of the TDSE. We implement the SCTS by solving Newton's equation of motion using a fourth-order Runge-Kutta method with adaptive step size \cite{Nuref}. In order to fully resolve the rich interference structure, we need to use the a momentum-space bin size of $\Delta k_x=0.0019$~a.u. For this value of $\Delta k_x$ the convergence of the interference oscillations is achieved for an ensemble consisting of $1.2\times10^{7}$ trajectories. At first, we consider photoelectron momentum distributions. In Fig.~2~(a) we compare the SCTS model with the solution of the TDSE. The TDSE photoelectron momentum distribution has a rather complicated structure. This is due to the fact that the laser pulse used in calculations is neither long nor very short. The side maxima at $k_x=-1.35$~a.u. and $k_x=1.33$~a.u. are created due to the interference of contributions from times near the central maximum and minimum of the vector potential, respectively, see Fig.~1(c). The central minimum of the vector potential is also responsible for the formation of the maximum at $k_x=1.0$~a.u. On the other hand, the ATI peaks in the electron momentum distributions are most pronounced in the range of $k_x$ from $-1.0$~a.u. to $-0.25$~a.u. The SCTS model predicts a caustic of the momentum distribution around $k_x=0.38$~a.u. For this reason, we normalize the distributions of Fig.~2~(a) to the total ionization yield. Fig.~2~(a) shows that there is only a qualitative agreement between the SCTS approach and the TDSE result. Indeed, the SCTS model underestimates the width of the momentum distribution.
\begin{figure}[h]
\begin{center}
\includegraphics[width=.6\textwidth]{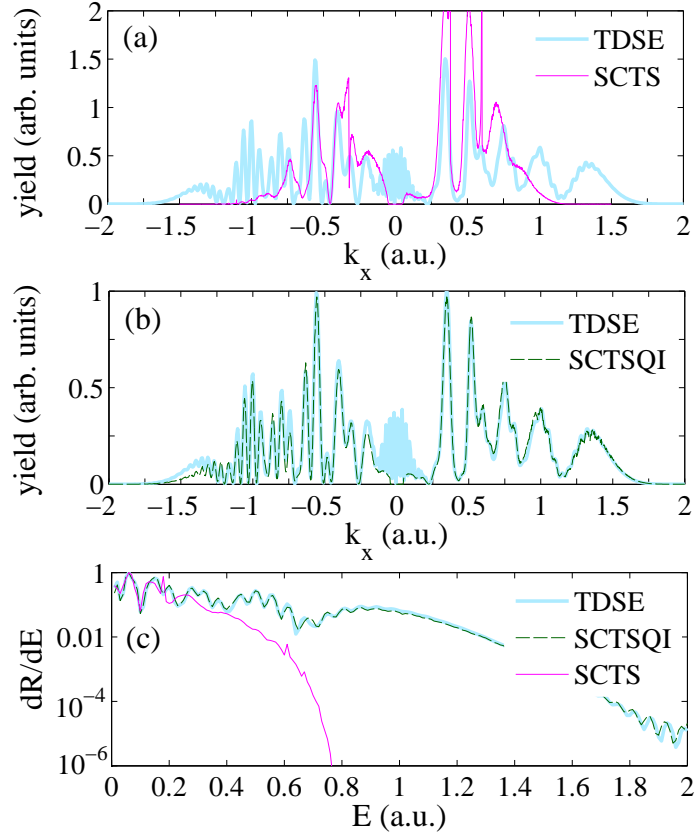} 
\end{center}
\caption{Comparison of the semiclassical models with the TDSE. The parameters are the same as in Fig.~1~(b). (a) The photoelectron momentum distributions for ionization of a one-dimensional model atom obtained from the SCTS model [magenta (thin) curve] and the solution of the TDSE [light blue (thick) curve]. The distributions are normalized to the total ionization yield. (b) The electron momentum distributions calculated using the present SCTSQI model [dark green (dashed) curve] and the TDSE [light blue (thick) curve]. The distributions are normalized to the peak values. (c) Electron energy spectra obtained from the TDSE [light blue (thick) curve], SCTSQI [dark green (dashed) curve], and the SCTS [magenta (thin) curve]. The spectra are normalized to the peak values.}
\label{fig2}
\end{figure} 

In Fig.~2~(b) we compare the SCTSQI model with the TDSE. In our SCTSQI simulations we have used $N=50$, $x_{\text{max}}=500$~a.u., and $x_b=70$~a.u. In order to achieve convergence of the momentum distribution, the bin size was chosen to be $1.5\times10^{-4}$~a.u., and $n_{p}=10^{6}$ trajectories were launched at every time step of the TDSE propagation. We note that in the mask method it is difficult to achieve full convergence of the TDSE momentum distribution for small momenta. The distribution in the vicinity of $k_x=0$ is formed by the slow parts of the electron wave packet. A long propagation time is needed, in order to let these parts reach the absorbing mask, and, therefore, to obtain converged distribution for small $k_x$. Thus we do not consider the region of small $k_x$ when comparing the SCTSQI with the TDSE. It is clearly seen from Fig.~2~(b) that for $\left|k_x\right|\gtrsim 0.15$~a.u. the SCTSQI model provides \textit{quantitative} agreement with fully quantum-mechanical result. This applies to both the width of the momentum distribution and the positions of the interference maxima (minima). The small remaining discrepancy in the heights of some of the interference maxima is caused by the fact that similar to the SCTS, the SCTSQI model does not account for the preexponential factor of the semiclassical matrix element \cite{Miller1974}.  

In Fig.~2~(c) we present the photoelecton energy spectra obtained from the SCTS, the solution of the TDSE, and the present SCTSQI model. It is seen that the SCTSQI and the TDSE spectra are almost identical, while the spectrum predicted by the SCTS model falls off to rapidly with the increase of the electron energy. This is a direct consequence of the fact that the SCTS model underestimates the width of the electron momentum distribution, see Fig.~2~(a).   

In order to further test the SCTSQI model, we calculate the electron momentum distributions for different positions of the mask $x_b$ and fixed $x_{\text{max}}$ of the computational box, see Fig.~3~(a). The distributions corresponding to different values of $x_b$ are in good quantitative agreement with each other. The same is also true for momentum distributions obtained for fixed $x_b$ and different values of $x_{\text{max}}$, see Fig.~3~(b). Here, we have used the two values $x_{\text{max}}=500$~a.u. and $x_{\text{max}}=200$ a.u. It should be stressed that it is impossible to obtain accurate electron momentum distributions for the small value $x_{\text{max}}=200$~a.u. using the mask method. We also note that for the 1D soft-core Coulomb potential used in this work, the smallest allowed $x_b$ should exceed 30-40~a.u., to be outside of the region where the bound-state wave function is localized. Indeed, due to the large number of time steps, even the absorption of a small fraction of the bound-state wave function at each step will result in a severe distortion of the final momentum distribution. 
\begin{figure}[h]
\begin{center}
\includegraphics[width=.8\textwidth]{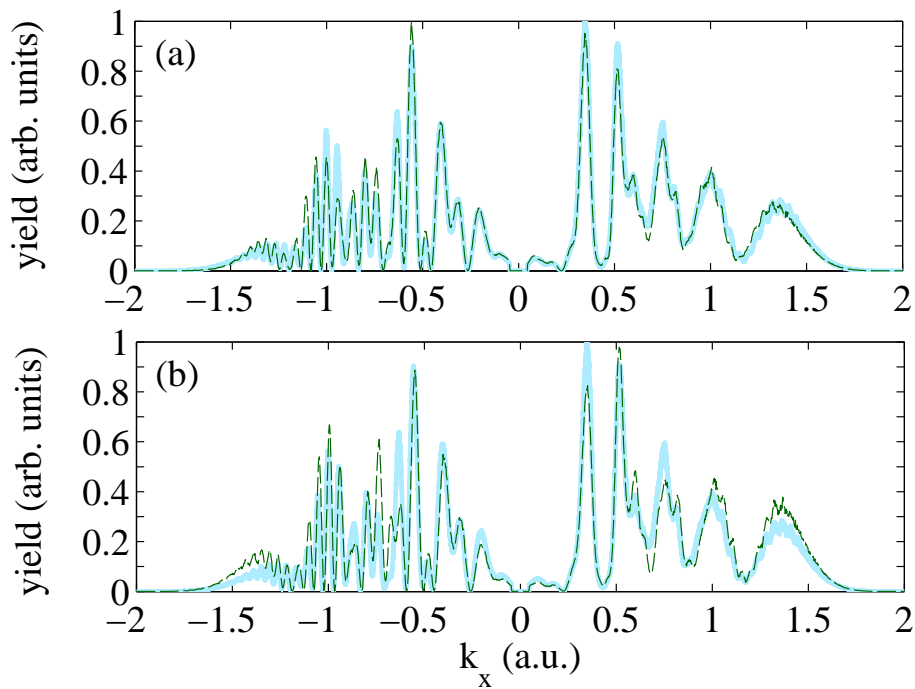} 
\end{center}
\caption{The outcomes of the SCTSQI model for different internal boundaries of the absorbing mask and lengths of the computational box. The distributions are normalized to the peak values. (a) The one-dimensional momentum distributions calculated within the SCTSQI model for the absorbing mask beginning at $x_b=50$~a.u. [light blue (thick) curve] and $x_b=100$~a.u. [dark green (dashed) curve]. The parameters are the same as in Fig.~1~(b), and the size of the computational box is $x_{\text{max}}=500$~a.u. (b) The one-dimensional momentum distributions obtained from SCTSQI for $x_{\text{max}}=500$~a.u. [light blue (thick) curve] and $x_{\text{max}}=200$~a.u. [dark green (dashed) curve]. The parameters are the same as in Fig.~1~(b). The absorbing mask begins at $x_b=50$~a.u.}
\label{fig2}
\end{figure} 
Finally, we check how important is the phase of the factor $G\left(x,p_x,t\right)$ in Eq.~(\ref{sctsqi}). To this end, in Fig.~4 we compare photoelectron momentum distribution calculated using the formula 
\begin{equation}
R\left(k_x\right)=\left|\sum_{m=1}^{N_{T}}\sum_{j=1}^{n_{k_{x}}}\left|G\left(t_{0}^{m},x_{0}^{j},p_{x,0}^{j}\right)\right|\exp\left[i\Phi_{0}\left(t_{0}^{m},x_{0}^{j},p_{x,0}^{j}\right)\right]\right|^{2},
\label{sctsqi_abs}
\end{equation}
instead of the Eq.~(\ref{sctsqi}). We find that neglecting the phase of the Gabor transform is severe: The SCTSQI distribution cannot even be qualitatively reproduced when using Eq.~(\ref{sctsqi_abs}). This result could be expected. Indeed, the factor $G\left(x,p_x,t\right)$ contains all the information about the quantum dynamics of the absorbed part of the wave packet \textit{prior} its conversion to the ensemble of classical trajectories. In a sense the $I_pt_0$ term in the SCTS phase [see Eq.~(\ref{Phi_sim_1d})] plays the role of the phase $G\left(t,x,p_x\right)$ of the Gabor transform in Eq.~(\ref{sctsqi}). 
\begin{figure}[h]
\begin{center}
\includegraphics[width=.55\textwidth]{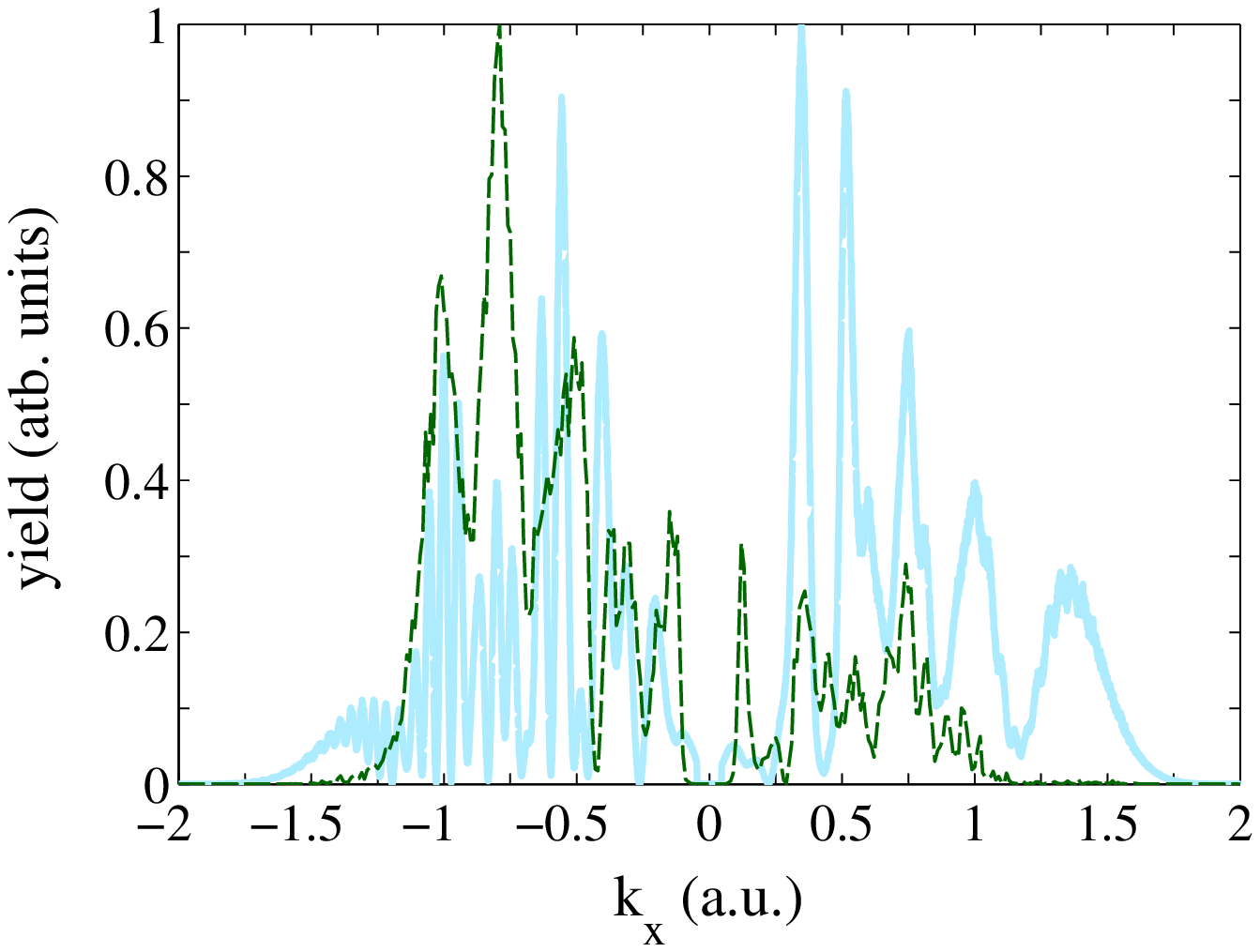} 
\end{center}
\caption{The photoelectron momentum distributions obtained from the SCTSQI model [light blue (thick) curve] and using Eq.~(\ref{sctsqi_abs}), i.e., neglecting the phase of the Gabor transform [dark green (dashed) curve]. The parameters are the same as in Figs.~1~(b), 2, and 3. The size of the computational box is $x_{\text{max}}=500$~a.u., and the absorbing mask begins at $x_b=50$~a.u. The distributions are normalized to the peak values.} 
\label{fig2}
\end{figure}

\section{Conclusions and outlook}

In conclusion, we have developed a trajectory-based approach to strong-field ionization: the semiclassical two-step model with quantum input. In the SCTSQI every trajectory is associated with the SCTS phase and, therefore, the SCTSQI model allows us to describe quantum interference and account for the ionic potential beyond the semiclassical perturbation theory. Furthermore, the SCTSQI corrects the inaccuracies of the SCTS model in treating the tunneling step. This has been achieved by the numerical solution of the TDSE with absorbing boundary conditions in a restricted area of space, applying the Gabor transform to the part of the wave function that is absorbed at each time step, and transforming this absorbed part into classical trajectories. The Gabor transform determines quantum amplitudes assigned to trajectories of the ensemble. Therefore, in the SCTSQI model the initial conditions of classical trajectories are governed by the exact quantum dynamics rather than by the quasistatic or SFA-based expressions as in other semiclassical approaches.   

We have tested our SCTSQI model by comparing its predictions with the numerical solution of the 1D TDSE. We have shown that the SCTSQI model yields quantitative agreement with the fully quantum results. This is true not only for the widths of the electron momentum distributions, but also for the positions of the interference maxima and minima. The model can be straightforwardly extended to the three-dimensional case. % Its numerical implementation can be made very efficient by using the sliding fast Fourier transform. 
Most importantly, the SCTSQI circumvents the non-trivial problem of choosing the initial conditions for classical trajectories. This makes the SCTSQI model extremely useful for study of strong-field ionization of molecules.   

\section{Acknowledgment} 
We are grateful to Professor Lars Bojer Madsen (Aarhus University), as well as to Nicolas Eicke and Simon Brennecke (Leibniz Universit\"{a}t Hannover) for stimulating discussions. This work was supported by the Deutsche Forschungsgemeinschaft (Grant No.~SH~1145/1-1).

\end{document}